\documentclass[12pt]{article}
\usepackage{amsfonts}
\usepackage{amsmath}
\usepackage{amssymb}
\usepackage{graphicx}
\usepackage{color}
\usepackage[all, knot]{xy}
\usepackage{tikz}

\usepackage[utf8]{inputenc}
\usepackage{epstopdf}
\usepackage[footnotesize]{caption}
\usepackage{amsthm}
\usepackage{enumitem}
\usepackage{mathrsfs}

\usepackage[margin=3cm]{geometry}

\def \be {\begin{equation}}
\def \ee {\end{equation}}
\def \bea {\begin{eqnarray}}
\def \eea {\end{eqnarray}}
\def \nn {\nonumber}

\def \rr {\raise.35ex\hbox{\small $\prime$}\kern-.17em{\mbox{\large $\imath$}}}

\def \dels {\partial\kern-.6em /\kern.1em}
\def \As {{A\kern-.5em / \kern.5em}}
\def \Ds {D\kern-.7em / \kern.5em}

\def \ks {k\kern-.5em /}
\def \ls {l\kern-.5em /}

\newcommand{\ci}[1]{}

\newcommand{\ba}{\begin{eqnarray}}
\newcommand{\ea}{\end{eqnarray}}
\newcommand{\bal}{\begin{align}}
\newcommand{\eal}{\end{align}}
\newcommand{\bay}[1]{\left(\begin{array}{#1}}
\newcommand{\eay}{\end{array}\right)}

\newcommand{\hide}[1]{}

\newlist{axioms}{enumerate}{2}
\setlist[axioms,1]{label=\textbf{A\arabic{axiomsi}.}, ref=A\arabic{axiomsi}}
\setlist[axioms,2]{label=\textbf{A\arabic{axiomsi}\rlap{\myEnumCounter{axiomsii}}.},%
                   ref=A\arabic{axiomsi}\myEnumCounter{axiomsii},%
                   align=parleft,%
                   leftmargin=0em,%
                   itemsep=1.4ex,%
                   before={\stepcounter{axiomsi}}}
                   
  \usetikzlibrary{decorations.markings}

\begin{document}

\begin{titlepage}

\begin{center}

\hfill
\vskip .2in

\textbf{\LARGE
Two-Dimensional Dilaton Gravity Theory
and Lattice Schwarzian Theory
\vskip.3cm
}
\vskip .5in
{\large
Su-Kuan Chu$^{a, b}$ \footnote{e-mail address: skchu@terpmail.umd.edu}, 
Chen-Te Ma$^{c, d, e}$ \footnote{e-mail address: yefgst@gmail.com},
and Chih-Hung Wu $^{e, f}$ \footnote{e-mail address: b02202007@ntu.edu.tw}
\\
\vskip 1mm
}
{\sl
$^a$
Joint Quantum Institute, NIST/University of Maryland, College Park,\\
 Maryland 20742, USA.
\\
$^b$
Joint Center for Quantum Information and Computer Science,\\
 NIST/University of Maryland, College Park, Maryland 20742, USA.
 \\
 $^c$
 Institute of Quantum Matter,\\
  School of Physics and Telecommunication Engineering,\\ 
South China Normal University, Guangzhou 510006, China.
\\
$^d$
The Laboratory for Quantum Gravity and Strings,\\
Department of Mathematics and Applied Mathematics,
University of Cape Town, Private Bag, Rondebosch 7700, South Africa.
\\
$^e$
Department of Physics and Center for Theoretical Sciences, \\
National Taiwan University, Taipei 10617, Taiwan, R.O.C..
\\
$^f$
Department of Physics, University of California Santa Barbara,
Broida Hall,\\
Bldg. 572,
Santa Barbara, CA 93106-9530, USA.
}\\
\vskip 1mm
\vspace{40pt}
\end{center}
\newpage
\begin{abstract}
We report a holographic study of a two-dimensional dilaton gravity theory with the Dirichlet boundary condition for the cases of non-vanishing and vanishing cosmological constants. Our result shows that the boundary theory of the two-dimensional dilaton gravity theory with the Dirichlet boundary condition for the case of non-vanishing cosmological constants is the Schwarzian term coupled to a dilaton field, while for the case of vanishing cosmological constant, a theory does not have a kinetic term. We also include the higher derivative term $R^2$, where $R$ is the scalar curvature that is coupled to a dilaton field. We find that the form of the boundary theory is not modified perturbatively. Finally, we show that a lattice holographic picture is realized up to the second-order perturbation of boundary cut-off $\epsilon^2$ under a constant boundary dilaton field and the non-vanishing cosmological constant by identifying the lattice spacing $a$ of a lattice Schwarzian theory with the boundary cut-off $\epsilon$ of the two-dimensional dilaton gravity theory.
\end{abstract}
\end{titlepage}

\section{Introduction}
\label{1}
The holographic principle \cite{Bousso:2002ju} states that the physical degrees of freedom of quantum gravity theory can be described by the boundary if the phenomenon can be observed outside the black hole \cite{tHooft:1993dmi}. The holographic principle was motivated by the area law of black hole entropy \cite{Strominger:1994tn}, which was found by generalizing the second law of ordinary thermodynamics, which describes that entropy does not decrease in a closed system. The generalized second law of thermodynamics \cite{Bekenstein:1972tm} is that the sum of black hole entropy and environmental entropy outside does not decrease. The coefficient of the black hole entropy was also first computed from the generalized second law of thermodynamics by assuming that the smallest possible radius of one particle is equal to its Compton wavelength \cite{Bekenstein:1973ur}. Nevertheless, it had already been known that the assumption for the radius is not suitable, and the assumption \cite{Bekenstein:1973ur} did not give the correct coefficient. After the four laws of black hole thermodynamics was proposed \cite{Bardeen:1973gs}, and the Hawking temperature was derived from the semi-classical approximation (the metric field is treated as a background field, but the matter field can have quantum fluctuation), the exact coefficient of the black hole entropy in the 3+1 dimensional Schwarzschild solution was obtained by using the first law of black hole thermodynamics and the Hawking temperature \cite{Hawking:1974sw}.
\\

\noindent 
The above discussion of the holographic principle only provides the evidence from Einstein gravity theory, which possibly is an infrared (IR) effective theory only. The connection between an ultraviolet (UV) complete theory and the holographic principle is still absent. The first evidence of the holographic principle from a UV complete theory was discovered from string theory by showing that the area of a system can be related to the string coupling constant \cite{Susskind:1994vu}. String theory is defined on the worldsheet space and its low-energy theory contains Einstein gravity theory from the fluctuation of target space and vanishing one-loop $\beta$-function. From the solutions of low-energy string theory or the type IIB supergravity theory, it was found that the bulk metric is the five-dimensional anti-de Sitter (AdS$_5$) times a five-dimensional sphere and the boundary theory is the four-dimensional ${\cal N}=4$ U($N$) supersymmetric Yang-Mills theory, where ${\cal N}$ is the number of supercharge \cite{Maldacena:1997re}. This also leads to the conjecture for the general correspondence between a theory in the $d$-dimensional anti-de Sitter spacetime and the ($d-1$)-dimensional conformal field theory (AdS$_d$/CFT$_{d-1}$) \cite{Maldacena:1997re}. 
\\

\noindent 
The interesting holographic two-dimensional bulk gravity theory \cite{Brown:1988am} is a dilaton gravity theory with the Dirichlet boundary condition and a negative cosmological constant \cite{Jackiw:1984je}. It was shown that the associated boundary action is the Schwarzian term \cite{Jensen:2016pah} coupled to a dilaton field, the Lyapunov exponent is also saturated in the boundary theory \cite{Jensen:2016pah}, and the effective boundary action is also equivalent to the one-dimensional Liouville theory \cite{Engelsoy:2016xyb}. Furthermore, the two-dimensional dilaton gravity theory can also be obtained from the three-dimensional Einstein gravity theory through compactification \cite{Achucarro:1993fd}. The non-trivial feature of the holographic study is that a generic dilaton solution \cite{Cadoni:1994uf} on the boundary can be consistently obtained from the bulk side \cite{Maldacena:2016upp}. The dilaton gravity theory also shares the solvable property from different approaches \cite{Saad:2019lba} \cite{Yang:2018gdb}. It was also shown that the boundary effective theory is Chern-Simons quantum mechanics by including a counter-term for the flat background when the boundary dilaton is a constant \cite{Dubovsky:2017cnj}.
\\

Now the holographic principle is still largely restricted to the AdS metric and the extension of other metrics based on the AdS metric \cite{Kitaev:2018wpr} \cite{Anninos:2018svg}. In this Letter, we show that the holographic study can be extended to other metric solutions at the classical level. We also show that the higher derivative term $R^2$ can be easily included in the two-dimensional dilaton gravity theory because after integrating out the dilaton field, the physical degrees of freedom of a two-dimensional gravity theory only comes from the scalar curvature ($R$).
\\

Additionally, a boundary cut-off is usually introduced in a bulk gravity theory to study the holographic correspondence \cite{Susskind:1998dq}. Because we expect that such cut-off would correspond to a UV cut-off in the corresponding boundary theory, and that a real space lattice spacing in a lattice theory should play the same role as the UV cut-off, an action of a two-dimensional dilaton lattice gravity theory possibly corresponds to the action of lattice boundary theory \cite{Stanford:2017thb}. The central interesting direction in this letter is: {\it Could we put the Schwarzian theory on a lattice with the exact symmetry of bulk AdS isometry?}  In this Letter, we use a perturbation method to show such equivalence up to the second-order of boundary cut-off $\epsilon^2$. This provides a deformation from the Schwarzian theory with a finite lattice effect. The similar study was studied by the ${\cal T}{\cal \bar{T}}$ deformation \cite{Dubovsky:2018bmo} \cite{Gross:2019ach}. Here the correction still preserves the SL(2) symmetry without breaking from the lattice effect. This is our advantage in this construction for the symmetry perspective.

\section{Holographic Study\\
in a Two-Dimensional Dilaton Gravity Theory}
\label{2}
A holographic study of a two-dimensional dilaton gravity theory with the Dirichlet boundary condition and a negative cosmological constant has been investigated \cite{Jensen:2016pah} \cite{Engelsoy:2016xyb} \cite{Maldacena:2016upp}. Besides reproducing their result, we also consider the scenario with positive and zero cosmological constants and take into account an effect of the higher derivative term $R^2$. 
\\

We first consider that the action ($S_{\mathrm{gravity}}$) of the two-dimensional gravity theory in the bulk as
\bea
S_{\mathrm{gravity}}&\equiv&-\frac{\phi_0}{16\pi G_2}\int d^2x\sqrt{|\det g_{\mu\nu}|}\ R
\nn\\
&&-\frac{1}{16\pi G_2}\int d^2x\sqrt{|\det{g_{\mu\nu}}|}\ \phi (R-2\Lambda),
\eea
where $\phi_0$ is a trivial dilaton field or a constant field, $G_2$ is a two-dimensional gravitational constant, $\phi$ is a non-trivial dilaton field. Note that the spacetime indices are labeled by the Greek indices ($\mu$, $\nu$, $\cdots$), and that the spacetime interval ($ds^2$) and the metric ($g_{\mu\nu}$) are defined by $ds^2\equiv g_{\mu\nu}dx^{\mu}dx^{\nu}$. The scalar curvature ($R$) is defined by $R\equiv g^{\mu\nu}R_{\mu\nu}$ and $R_{\mu\nu}\equiv\partial_{\delta}\Gamma^{\delta}_{\nu\mu}-\partial_{\nu}\Gamma^{\delta}_{\delta\mu}
+\Gamma^{\delta}_{\delta\lambda}\Gamma^{\lambda}_{\nu\mu}
-\Gamma^{\delta}_{\nu\lambda}\Gamma^{\lambda}_{\delta\mu}$, where
 $\Gamma^{\mu}_{\nu\delta}\equiv (1/2)g^{\mu\lambda}\big(\partial_{\delta}g_{\lambda\nu}+\partial_{\nu}g_{\lambda\delta}
-\partial_{\lambda}g_{\nu\delta}\big)$.
\\

Since the first term in the action is also topological, the topological term contributes the trivial equation of motion. Nevertheless, the second term of the action gives the equation of motion $R=2\Lambda$ by varying with respect to the non-trivial dilaton field $\phi$. The  equation of motion of the inverse metric field $g^{\rho\sigma}$ is $-\nabla_{\rho}\nabla_{\sigma}\phi+g_{\rho\sigma}\nabla^2\phi+\Lambda g_{\rho\sigma}\phi=0$, where $\nabla_{\rho}$ is the covariant derivative of $\rho$ direction, and $\nabla_{\sigma}$ is the covariant derivative of $\sigma$ direction, and $\nabla^2\equiv g^{\mu\nu}\nabla_{\mu}\nabla_{\nu}$ is the Laplace operator. 
\\

When the Dirichlet boundary condition is imposed on the boundary fields: $\delta  g_{\mu\nu}=0$ and $\delta \phi=0$, one automatically includes
the boundary term in the action of two-dimensional dilaton gravity theory $-\lbrack1/(8\pi G_2)\rbrack\int du\sqrt{|\det{h_{uu}}|}\ \phi K$, where $h_{uu}$ is an induced metric, and $K$ is the trace of the extrinsic curvature $K\equiv g^{\mu\nu}\nabla_{\nu}n_{\mu}$, in which the unit vector $n_{\mu}$ satisfies the normalization condition: $g_{\mu\nu}n^{\mu}n^{\nu}=-1$ for $\Lambda>0$ and $g_{\mu\nu}n^{\mu}n^{\nu}=1$ for $\Lambda\le0$ and $\nabla_{\nu}n_{\mu}\equiv\partial_{\nu}n_{\mu}-\Gamma^{\gamma}_{\mu\nu}n_{\gamma}$. Because the variation of the $R^2$ term gives the $R\delta R$, and the variation of the dilaton field gives the constant scalar curvature, the boundary term is proportional to the $K$, and the coefficient contains the constant scalar curvature.
\\

For the case of the non-vanishing cosmological constant, we consider the metric $ds^2_{2dw}\equiv-(1/\Lambda)\lbrack(dt^2+dz^2)/z^2\rbrack$ and the induced metric $h_{uu}=-(1/\Lambda)\lbrack(t^{\prime 2}+z^{\prime 2})/z^2\rbrack$, where $t^{\prime}\equiv dt/du$ and $z^{\prime}\equiv dz/du$, and also use the proper length of the induced metric at leading order with respect to the boundary cut-off $(1/|\Lambda|)\lbrack(t^{\prime 2}+z^{\prime 2})/z^2\rbrack=1/\epsilon^2+\cdots$, where $\epsilon$ is a positive constant. The perturbative solution of $z$ is $z=\big(\epsilon/\sqrt{|\Lambda|}\big) t^{\prime}+\cdots$. When the cosmological constant vanishes, we choose the spacetime interval $ds^2_{2dn}\equiv dt^2+dx^2$ and the induced metric $h_{uu}=t^{\prime 2}+x^{\prime 2}=1/\epsilon^2$. We can also parametrize $t^{\prime}$ and $x^{\prime}$ as $t^{\prime}=(1/\epsilon)\cos\theta$ and $x^{\prime}=(1/\epsilon)\sin\theta$, where $0\le\theta < 2\pi$.
\\

To derive the boundary action of the two-dimensional dilaton gravity theory with the Dirichlet boundary condition, we first integrate out the bulk dilaton field ($\phi$), and then we find that the action of the bulk theory vanishes and only need to consider the boundary action of the two-dimensional dilaton gravity theory with the constraint $R=2\Lambda$. Therefore, we take solutions under the constraint ($R=2\Lambda$) to the boundary action of the two-dimensional dilaton gravity theory with the Dirichlet boundary condition. 
\\

For the case of the non-vanishing cosmological constant, we find that the boundary action of the two-dimensional dilaton gravity theory with the Dirichlet boundary condition is 
\bea
&&-\frac{1}{8\pi G_2}\int du \sqrt{|h_{uu}|}\ \phi K
\nn\\
&=&
-\frac{1}{8\pi G_2}\int du\ \phi_b Sch(t, u)+\cdots,
\eea
in which $\cdots$ is not universal with respect to the boundary cut-off $\epsilon$, the Schwarzian term is $Sch(t, u)\equiv (t^{\prime\prime\prime}/t^{\prime})-(3/2)(t^{\prime\prime 2}/t^{\prime 2})$, 
and the dilaton field on the boundary is $\phi=|\Lambda|\phi_b/\epsilon+\cdots$ \cite{Cadoni:1994uf} \cite{Maldacena:2016upp}, where 
$\phi_b\equiv (\alpha+ \beta t+\gamma t^2)/t^{\prime}$, in which $\alpha$, $\beta$, and $\gamma$ are arbitrary constants. Here we assume that the boundary cut-off $\epsilon$ is small enough for performing a valid perturbation. We can get the same boundary action for the cases of the positive and the negative cosmological constants because the normalization condition of unit vector $n_{\mu}$ has an opposite sign.
\\

For the case of vanishing cosmological constant, we find that the corresponding boundary action is 
\bea
-\frac{1}{8\pi G_2}\int du \sqrt{|h_{uu}|}\ \phi K
=-\frac{1}{8\pi G_2}\int du\ \phi_b \theta^{\prime}.
\eea
 The dilaton field on the boundary is $\phi=\phi_b$ \cite{Dubovsky:2017cnj} because the solution of the dilaton field in the bulk is $\phi=\alpha t+\beta x+\gamma$, where $\alpha$, $\beta$, and $\gamma$ are again arbitrary constants. The notation $\theta$ is a $2\pi$ periodic field. Hence the boundary theory of two-dimensional dilaton gravity theory with the Dirichlet boundary condition and the vanishing cosmological constant loses a kinetic term. Because of our work focus on checking the consistent equations of motion from the bulk and boundary sides, we do not put a counter-term \cite{Dubovsky:2017cnj}. At the quantum level, the counter-term is necessary.
\\

For the case of non-vanishing cosmological constants, the variation with respect to the boundary field $t$ results in the consistent solution of boundary dilaton field as in the bulk side \cite{Jensen:2016pah} \cite{Engelsoy:2016xyb} \cite{Cadoni:1994uf} \cite{Maldacena:2016upp}. This means that we can use the boundary field $t$ to be the measure in the boundary theory. We can also perform the variation with respect to the boundary field $\theta$ to get the constant solution of boundary dilaton field, which can also be obtained from the bulk side, for the case of vanishing cosmological constant. Here the boundary field $\theta$ can also be served as the measure in the corresponding boundary theory. 
\\

Now we include the higher derivative term $R^2$, which is coupled to the non-trivial dilaton field ($\phi$) in the bulk theory, and derive the corresponding boundary theory. We begin with the action
\bea
&&S_{\mathrm{gravity},bR^2}
\nn\\
&\equiv&-\frac{\phi_0}{16\pi G_2}\int d^2x\sqrt{|\det g_{\mu\nu}|}\ R
\nn\\
&&-\frac{1}{16\pi G_2}\int d^2x\sqrt{|\det{g_{\mu\nu}}|}\ 
\phi (R-2\Lambda+b R^2),
\eea 
where $b$ is a constant. The first term of the action ($S_{\mathrm{gravity},bR^2}$) can also be ignored since it is a topological term. Even if we include the higher derivative term $R^2$, the model is still exactly solvable. Integrating out the non-trivial dilaton field gives rise to the constraint $b R^2+R-2\Lambda=0$. There are two solutions with the same cosmological constant. Because we are considering a perturbative effect, we require that taking the limit $b\rightarrow 0$ recovers $R\rightarrow 2\Lambda$.
Therefore, we choose the solution for the scalar curvature to be $R=(-1+\sqrt{1+8b\Lambda})/(2b)$.
\\

Because we also put the Dirichlet boundary condition on the boundary fields, a boundary theory comes from the boundary term $-\lbrack1/(8\pi G_2)\rbrack\int du\sqrt{|h_{uu}|}\ \phi(1+2b R)K$. Therefore, we can say that the form of the boundary action should not be modified if the dilaton solution is not modified.
\\

Because we only consider the leading correction of the higher derivative term $R^2$, we need to carry out a perturbation analysis with respect to the parameter $b$. We perform the variation with respect to the inverse metric field $g^{\rho\sigma}$ and find that the solution of the non-trivial dilaton field ($\phi$) is followed by $-\nabla_{\rho}\nabla_{\sigma}\phi+g_{\rho\sigma}\nabla^2\phi+\tilde{\Lambda} g_{\rho\sigma}\phi+\cdots=0$, in which $\cdots$ cannot be controlled by the higher derivative term $R^2$ and $\tilde{\Lambda}\equiv \Lambda(1-2b\Lambda)$. Hence the solution of the non-trivial dilaton field ($\phi$) is not modified by including the higher derivative term $R^2$.
\\

When we do the generalization to the local higher derivative terms $f(R)$, the scalar curvature is still a constant after integrating out the dilaton field. This immediately implies that boundary action is still the Schwarzian theory, but the consistent solution of the dilaton field on the boundary needs to be checked order-by-order.
\\

In this section, we showed that the Schwarzian theory is quite generic even if we include the higher derivative term. This means that the generalization to the lattice does not lose the generality. We will construct the lattice Schwarzian theory in the next section.

\section{Two-Dimensional Dilaton Gravity Theory\\
and Lattice Schwarzian Theory}
\label{3}
Now we generalize the holographic duality from the continuum theory to lattice theory with a finite lattice spacing and lattice size. We propose a lattice Schwarzian theory with an SL(2) symmetry. The lattice Schwarzian theory extends the SL(2) symmetry to the bulk, not only on the boundary. This gives the different lattice correction than the Ref. \cite{Stanford:2017thb} because the lattice theory in the Ref. \cite{Stanford:2017thb} only requires the SL(2) symmetry on the boundary.
\\

The action of the lattice Schwarizian theory is
\bea
&&S_{\mathrm{lattice}}
\nn\\
&\equiv&\frac{1}{128\pi G_2a}\sum_{j=1}^n\bigg\lbrack\bigg(\frac{(f^*_{j+6}-f^*_{j+2})(f^*_{j+4}-f^*_j)}{(f^*_{j+6}-f^*_{j+4})(f^*_{j+2}-f^*_j)}-4\bigg)
\nn\\
&&+\bigg(\frac{(f_{j+6}-f_{j+2})(f_{j+4}-f_j)}{(f_{j+6}-f_{j+4})(f_{j+2}-f_j)}-4\bigg)\bigg\rbrack, 
\eea
and the measure of the lattice Schwarzian theory is 
\bea
&&\prod_{j=1}^n\frac{df_jdf_j^*}{(f_{j+1}-f_{j})(f^*_{j+1}-f_j^*)}
\nn\\
&&\times\delta\bigg(\frac{(f^*_{j+1}-f_{j-1}^*)(f_{j+1}-f_{j-1})}{(f_{j+1}-f_{j+1}^*)(f_{j-1}-f_{j-1}^*)}+\beta^2\bigg),
\eea
where $f_j\equiv f(u_j)\equiv t(u_j)+i z(u_j)$ is a complex field with $u_j=ja$ and $n$ is a number of lattice points in the $u$ direction. The notation $\beta^2$ is a parameter of this theory to be tuned to match a discretized boundary theory of the two-dimensional dilaton gravity theory. We choose the measure because we want to obtain the similar constraint to the bulk coordinates, $z$ and $t$, as in the two-dimensional dilaton gravity theory. Therefore, the formulation can be similar between the bulk and boundary theories.
\\

The lattice Schwarzian theory is invariant under the SL(2) transformation:
$f_i\rightarrow(\bar{a} f_i+\bar{b})/(\bar{c} f_i+\bar{d})$ and $\bar{a}\bar{d}-\bar{b}\bar{c}=1$, where $\bar{a}$, $\bar{b}$, $\bar{c}$ and $\bar{d}$ are real constants. Because the boundary theory of two-dimensional dilaton gravity theory is invariant under the SL(2) symmetry of $t+iz$, not the SL(2) symmetry of $t$, we choose the complex field $f=t+iz$ rather than choosing a real field $f=t$ to ensure that the symmetry of lattice Schwarzian theory is the same as the symmetry of the boundary term of the two-dimensional dilaton gravity theory. When one only considers the leading order of the lattice spacing or boundary cut-off, one only needs the SL(2) symmetry of $t$ since $z$ is at order $\epsilon$. For the next leading order, the symmetry of the discretized boundary theory of two-dimensional dilaton gravity theory or lattice Schwarzian theory should be the full SL(2) symmetry of $t+iz$. Because the lattice theory has the same symmetry as in the bulk, one should expect that the lattice theory can be valid at the next non-trivial leading order with respect to the lattice spacing, not only the non-trivial leading order. Hence the symmetry extension provides the complex fields in this lattice Schwarzian theory.
\\

To relate the action of the lattice Schwarzian theory to a boundary action of the two-dimensional dilaton lattice gravity theory with the Dirichlet boundary condition, the constant boundary dilaton field and the non-vanishing cosmological constant, we choose the relation between the lattice spacing $a$ and the boundary cut-off $\epsilon$ as $a=\beta\epsilon/\sqrt{|\Lambda|}$, where $\beta$ is the same parameter that we mentioned above.
\\

It is interesting to note that the constraint 
\bea
\delta\bigg(\frac{(f^*_{j+1}-f_{j-1}^*)(f_{j+1}-f_{j-1})}{(f_{j+1}-f_{j+1}^*)(f_{j-1}-f_{j-1}^*)}+\beta^2\bigg)
\eea
 in the measure gives
\bea
\frac{t^{\prime 2}+z^{\prime 2}}{z^2}&=&\frac{|\Lambda|}{\epsilon^2}
\nn\\
&&+\beta^2\frac{\epsilon^2}{|\Lambda|}
\nn\\
&&\times\bigg(-\frac{t^{\prime 2}z^{\prime 2}}{z^4}-\frac{z^{\prime 4}}{z^4}-\frac{1}{3}\frac{t^{\prime}t^{\prime\prime\prime}}{z^2}
\nn\\
&&-\frac{1}{3}\frac{z^{\prime}z^{\prime\prime\prime}}{z^2}+\frac{t^{\prime 2}z^{\prime\prime}}{z^3}+\frac{z^{\prime 2}z^{\prime\prime}}{z^3}\bigg)+\cdots
\eea
through the perturbation method, in which we used $f_j=f=t+iz$ and $f_{j+\alpha}=f+\alpha af^{\prime}+(\alpha^2/2)a^2f^{\prime\prime}+(\alpha^3/6)a^3f^{\prime\prime\prime}
+(\alpha^4/24)a^4f^{\prime\prime\prime\prime}+(\alpha^5/120)a^5f^{\prime\prime\prime\prime\prime}+\cdots$. This constraint is suggesting that the induced metric ($h_{uu}$) should be modified from the higher-order of the boundary cut-off in the two-dimensional dilaton gravity theory with the Dirichlet boundary condition. The perturbative solution of $z$ is $z=\bigg(\epsilon/\big(\sqrt{|\Lambda|}\big)\bigg) t^{\prime}+\bigg(\epsilon^3/|\Lambda|^{\frac{3}{2}}\bigg)\bigg(\big((\beta^2+1)/2\big)(t^{\prime\prime 2}/t^{\prime 2})-\big(\beta^2/3\big)(t^{\prime\prime\prime}/t^{\prime})\bigg)+\cdots$ by solving the constraint.
\\

On the gravity side, we use the following boundary term to obtain the boundary action of two-dimensional dilaton gravity theory with the Dirichlet boundary condition and the constant boundary dilaton field
\bea
-\frac{1}{8\pi G_2\epsilon}\int du\sqrt{|h_{uu}|}\ \bigg(K-\sqrt{|\Lambda|}\bigg),
\eea
in which we choose $1/\epsilon$ for the boundary dilaton field. Hence with the perturbative solution of $z$, this boundary action becomes
\bea
\label{bgt}
&&-\frac{1}{8\pi G_2\epsilon}\int du\sqrt{|h_{uu}|}\ \bigg(K-\sqrt{|\Lambda|}\bigg)
\nn\\
&=&\frac{1}{8\pi G_2}\int du\ \Bigg\lbrack\bigg(\frac{3}{2}\frac{t^{\prime\prime 2}}{t^{\prime 2}}-\frac{t^{\prime\prime\prime}}{t^{\prime}}\bigg)
\nn\\
&&+\frac{\epsilon^2}{|\Lambda|}\bigg\lbrack\frac{\beta^2}{3}\frac{t^{\prime\prime\prime 2}}{t^{\prime 2}}+\bigg(\frac{1}{8}-\frac{\beta^2}{4}\bigg)\frac{t^{\prime\prime 4}}{t^{\prime 4}}\bigg\rbrack\Bigg\rbrack+\cdots
\eea
up to a total derivative term.
\\

When we choose the parameter $\beta^2=3/2$, we can first do the partial integration by part  and replace the integration in \eqref{bgt} by the discretized sum
\bea
   \int du\rightarrow\sum_{j=1}^na,
\eea
we can identify the action of the lattice Schwarzian theory with the discretized boundary action of the two-dimensional dilaton gravity theory under the Dirichlet boundary condition, the constant boundary dilaton field and the non-vanishing cosmological constant up to the second-order of the boundary cut-off $\epsilon^2$. Hence the lattice Schwarzian theory and the bulk gravity theory give the same result.
\\

 Note that the parameter $\beta^2=3/2$ is the only choice for obtaining the consistent result between the discretized boundary theory of two-dimensional dilaton gravity theory and lattice Schwarzian theory. This choice shows that the square root of the induced metric $\sqrt{h_{uu}}$ will give yet another the Schwarzian term at the next order. Therefore, our calculation showed that other values of the parameter $\beta^2$ cannot lead to a consistent study while preserving the SL(2) symmetry. This is the indication that this choice is protected by the SL(2) symmetry.

\section{Outlook}
\label{4}
\noindent 
We showed the holographic study for the generic metric solutions in the two-dimensional dilaton gravity theory at the classical level and extended the holographic study to the lattice level. Putting gravity theory on a lattice should lose diffeomorphism. Our study implies that the isometry of the metric should play a more important role than diffeomorphism in a holographic study. Now we discuss the renormalization group flow at the lattice level. If we put the lattice Schwarzian theory on the boundary of gravity theory, we can integrate out some sites of the periodic bosonic field to apply the renormalization group flow to obtain an IR effective theory. According to the common lore of the AdS/CFT correspondence, the direction of the renormalization group flow of quantum theory should correspond to the inward radial direction of quantum gravity theory. Thus, a UV gravity theory is expected to be obtained from the renormalization group flow from the lattice Schwarzian theory in the continuum limit. This holographic correspondence was obtained from integrating out the bulk dilaton field. Therefore, we do not expect that the IR effective theory can provide the information to the bulk dilaton field. 

\section*{Acknowledgments}
We would like to thank Aleksey Cherman, Bei-Lok Hu, Xing Huang, Thomas G. Mertens, and Shinsei Ryu for their useful discussion. 
Su-Kuan Chu was supported by AFOSR, NSF QIS, ARL CDQI, ARO MURI, ARO and NSF PFC at JQI. 
Chen-Te Ma was supported by the Post-Doctoral International Exchange Program and China Postdoctoral Science Foundation, Postdoctoral General Funding: Second Class (Grant No. 2019M652926), and would like to thank Nan-Peng Ma for his encouragement. We would like to thank the National Tsing Hua University, Tohoku University, Okinawa Institute of Science and Technology Graduate University, Yukawa Institute for Theoretical Physics at the Kyoto University, Istituto Nazionale Di Fisica Nucleare - Sezione di Napoli at the Università degli Studi di Napoli Federico II, Kadanoff Center for Theoretical Physics at the University of Chicago, Stanford Institute for Theoretical Physics at the Stanford University, Kavil Institute for Theoretical Physics at the University of California of the Santa Barbara, National Tsing Hua University, Israel Institute for Advanced Studies at the Hebrew University of Jerusalem, Jinan University, Institute of Physics at the University of Amsterdam, Shing-Tung Yao Center at the Southeast University, and Institute of Theoretical Physics at the Chinese Academy of Sciences. 
Discussions during the workshops, ``Novel Quantum States in Condensed Matter 2017'', ``The NCTS workshop on correlated quantum many-body systems: from topology to quantum criticality'', ``String-Math 2018'', ``Strings 2018'', ``New Frontiers in String Theory'', ``Strings and Fields 2018'', ``Order from Chaos'', ``NCTS Annual Theory Meeting 2018: Particles, Cosmology and Strings'', ``The 36th Jerusalem Winter School in Theoretical Physics - Recent Progress in Quantum Field / String Theory'', ``Jinan University Gravitational Frontier Seminar'', ``Quantum Information and String Theory'', ``Strings 2019'', ``Amsterdam Summer Workshop on String Theory'', ``Youth Symposium on Theoretical High Energy Physics in Southeast University'', and ``Workshop on Holography and Quantum Matter'', were useful to complete this work.

\end{document}